**Science and its significant other: Representing the humanities in bibliometric scholarship**

Thomas Franssen & Paul Wouters (CWTS, Leiden University, The Netherlands)

1. <u>introduction</u>

Bibliometrics offers a particular representation of science (Wouters, 1999; Nicolaisen 2007). Through bibliometric methods a bibliometrician will always highlight particular elements of publications, and through these elements operationalize particular representations of science, while obscuring other possible representations from view. Understanding bibliometrics as representation implies that a bibliometric analysis is always performative; a bibliometric analysis brings a particular representation of science into being that potentially influences the science system itself (e.g. Wyatt et al., 2017). The performative effects of bibliometrics have been studied primarily in relation to individual researchers' behavior and how pervasive representations (and particular indicators) might influence this (De Rijcke et al., 2016). How bibliometrics influence the ways we think about, compare and contrast different scientific domains in general has however not been systematically analyzed. The pervasiveness of bibliometric representations of science in the contemporary science system warrants such a study. Moreover, a systematic, historical view of the development of bibliometrics might also offer this scientific community a better understanding of itself as well as the future of the discipline.

We are in particular interested in the ways the humanities have been represented throughout the history of bibliometrics, often in comparison to other scientific domains or to a general notion of 'the sciences'. Earlier reviews of bibliometric literature pertaining to the humanities exist (Nederhof, 2006; also part 2.3 in Moed, 2006; Huang & Chang, 2008; Ardanuy, 2013) but have been predominantly methodological in nature. They ask what bibliometric methods are suitable to use for research evaluation in the humanities (and social sciences) but do not engage with the question of representation.

There are three elements of each bibliometric representation of the humanities that we analyze in this paper. The first is the theoretical concept around which a representation is developed, such as 'scientificness' or 'internationality'. We also aim to understand part of the theoretical hinterland (Law 2004) in which the concept is embedded. The second is the bibliometric method that is developed to operationalize the theoretical concept. The third element is the data source(s) used to empirically develop the bibliometric method. We understand combinations of a concept, bibliometric method and data source as the configuration of the bibliometric system, inspired by the work of Rheinberger (2010; see also Wouters, 2006) on 'experimental systems'. We argue that each configuration, that operationalizes a (new) concept in a particular way, offers the possibility of a new bibliometric representation of the humanities. Building bibliometric data infrastructures is highly complex and the transformation of a particular element, such as a reference, into a bibliometric element takes significant investment (e.g. Wouters, 1999). By analyzing the combination of concepts, methods and data sources we aim to highlight the practical work that is necessary for bibliometric representations of science to come into being. This also explains the path-dependency of bibliometric centers that often remain loyal to particular concepts, methods and especially data sources for a long period of time even when these become highly contested.

Our review discusses bibliometric scholarship between 1965 and 2016 that studies the humanities empirically. We distinguish between two periods of bibliometric scholarship. The first period, between 1965 and 1989, is characterized by a sociological theoretical framework, the development and use of the Price index, and small samples of journal publications as data sources. In this period we distinguish between two configurations of the bibliometric system that are closely linked. The second period, from the mid-1980s up until the present day, is characterized by a new hinterland, that of science policy and research evaluation, in which bibliometric methods become embedded. New bibliometric methods, such as publication profiles and citation impact analysis, are developed and explored in the humanities and new data sources such as the WoS databases, annual reports, surveys and, recently, local and national publication databases are developed. We distinguish between a third and fourth configuration of the bibliometric system in this period based on the data sources developed.

## 2. Data and methods

We have collected publications that employ bibliometric methods in the humanities published between 1965 and 2016. We have refrained from defining 'the humanities' ourselves. As we are studying the representation of the humanities in bibliometrics, it is important for us that the author defines a certain corpus as (part of) the humanities. Moreover, as we are studying the use of bibliometric methods, we have only selected publications that contain an empirical analysis. To limit our analysis, we have not included bibliometric studies using altmetric data sources and have chosen not to discuss the field of science mapping in which the humanities are increasingly topic of research (see Colavizza, 2017 for an overview).

Our search strategy has combined various methods. We draw on earlier attempts to collect all publications in bibliometrics pertaining to the humanities, notably from Ardanuy (2013) and Nederhof (2006). We searched using various scholarly databases and have used CitNetExplorer (Van Eck & Waltman, 2014) to explore the citation network of collected publications to find missing links. Our aim was to collect the most important publications in which authors employ bibliometric methods with the aim to analyze, often comparatively, (particular aspects of) the humanities or humanities disciplines. To establish the relevance of publications we have only included those that have at least one relation (either as a citer or as being cited) with another publication in the network. We have refrained from a further analysis of the impact of these publications although this does vary widely.

In total we have selected 47 publications (see appendix 1 for the full list). We read all publications, and coded them for the type of data source they use in their empirical analysis, and whether they entail a comparison (e.g. between departments, disciplines, domains and/or countries) (table 1). We have subsequently manually clustered these publications and visualized their citation relations using CitNetExplorer (figure 1).

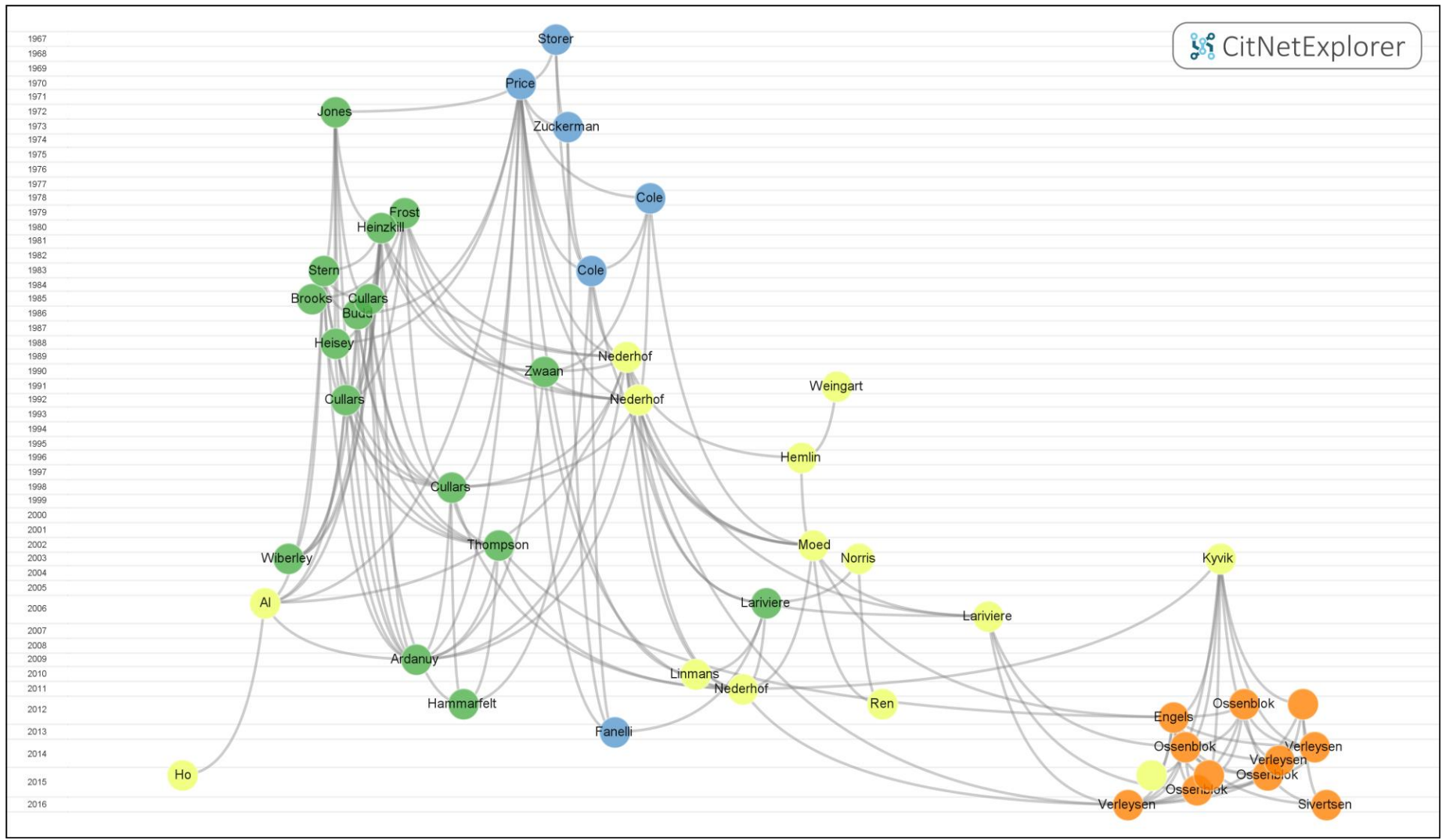

Figure 1: Visualization of full citation network 1965-2016

| Author(s) | Year | Reference lists | WoS databases | Citation impact | Publication profiles of departments/universities | Publication profile of disciplines/domains/countries |
|---|---|---|---|---|---|---|
| Storer | 1967 | X | | | | |
| Price | 1970 | X | | | | |
| Zuckerman & Merton | 1973 | X | | | | |
| Cole, Cole & Dietrich | 1978 | X | | | | |
| Cole | 1983 | X | | | | |
| Fanelli & Glanzel | 2013 | X | X | | | |
| Jones | 1972 | X | | | | |
| Frost | 1979 | X | | | | |
| Heinzkill | 1980 | X | | | | |
| Stern | 1983 | X | X | | | |
| Brooks | 1985 | X | | | | |
| Cullars | 1985 | X | | | | |
| Budd | 1986 | X | | | | |
| Heisey | 1988 | X | | | | |
| Zwaan & Nederhof | 1990 | X | X | | | |
| Cullars | 1992 | X | | | | |
| Cullars | 1998 | X | | | | |
| Thompson | 2002 | X | X | | | |
| Wiberley | 2003 | X | | | | |
| Larivière, Archambault, Gingras & Vignola-Gagné | 2006 | X | X | | | |
| Ardanuy, Urbano & Quintana | 2009 | X | | | | |
| Hammarfelt | 2012 | X | | | | |
| Nederhof, Zwaan, de Bruin & Dekker | 1989 | | X | X | X | |
| Nederhof & Noyons | 1992 | | X | X | X | |
| Weingart, Prinz, Kastner, Maasen & Walter | 1991 | | | | | X |
| Hemlin | 1996 | | | | X | |
| Moed, Nederhof & Luwel | 2002 | | | | X | |
| Norris & Oppenheim | 2003 | | X | X | X | |
| Kyvik | 2003 | | | | | X |
| Larivière, Gingras, Archambault | 2006 | | X | | | X |
| Al, Şahiner, Tonta | 2006 | X | X | X | X | X |
| Linmans | 2010 | | X | X | X | |
| Nederhof | 2011 | | X | X | X | |
| Ren & Gong | 2012 | | | X | | X |
| Ho & Ho | 2015 | | X | X | | X |
| Chinchilla-Rodriguez, Miguel & de Moya-Anegón | 2015 | | | X | | X |
| Engels, Ossenblok & Spruyt | 2012 | | | | | X |
| Sivertsen & Larsen | 2012 | | | | | X |
| Ossenblok, Engels & Sivertsen | 2012 | | | | | X |
| Verleysen & Engels | 2014 | | | | | X |
| Verleysen & Engels | 2014 | | | | | X |
| Ossenblok, Verleysen & Engels | 2014 | | | | | X |
| Ossenblok & Engels | 2015 | | | | | X |
| Hammarfelt & De Rijcke | 2015 | | | | X | |
| Ossenblok, Guns & Thelwall | 2015 | | | | | X |
| Sivertsen | 2016 | | | | | X |
| Verleysen & Weeren | 2016 | | | | | X |

Table 1: overview of selected publications and their data source clustered in four configurations of the bibliometric system.

3. The first period of bibliometric scholarship: an empirical hierarchy of sciences in sociology and librarians who study the humanities

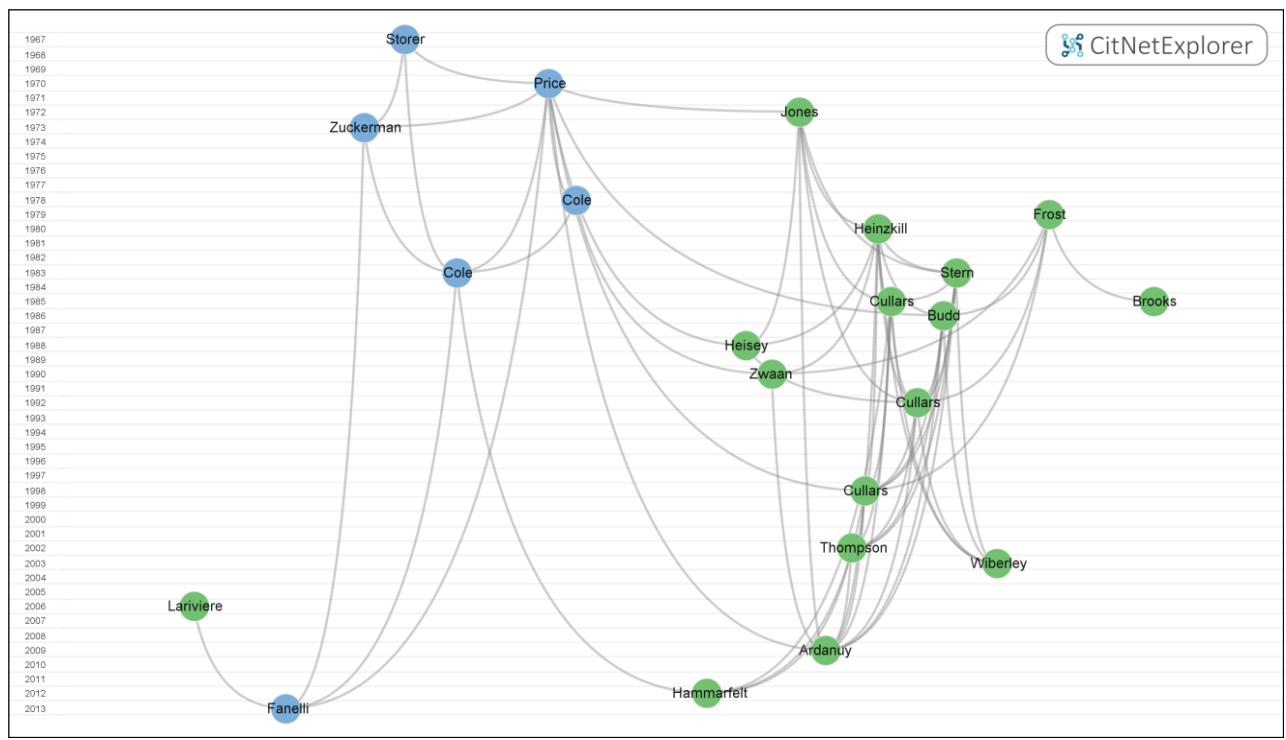

Figure 2: Visualization of the citation network of the first and second configuration of the bibliometric system

The first configuration of the bibliometric system was developed by early bibliometricians and sociologists in the 1960s and 1970s (see Godin, 2005; Cronin 1984; Garfield, Malin & Small, 1978; Narin, 1976). In these first two decades, bibliometric research was primarily based upon data extracted from the Science Citation Index (SCI). As humanities journals were not included in the SCI, with the exception of history and philosophy of science journals, bibliometric research that included the humanities was always based on smaller samples of bibliometric information gathered manually. Overall, very little attention was paid to the humanities in these early bibliometric studies. Only four publications included the humanities empirically and discussed them in some detail (the study on the humanities presented in Cole, 1983, is identical to the one in Cole, Cole & Dietrich, 1978). In them we find a common theoretical framework, bibliometric method and data sources. In all of these studies characteristics of publications (predominantly of the reference list) are used to analyze the 'scientificness', 'hardness' or 'codification' of scientific domains.

In these studies, the humanities became enacted through a particular notion of what constitutes scientificness, and thus in a particular hierarchy of the sciences. The four empirical studies we will briefly discuss are Storer's 'The hard sciences and the soft: some sociological observations' (1967), Price's 'citation measures of hard science, soft science, technology and nonscience' (1970), followed by Zuckerman and Merton's 'Age, aging and age structure in science' (1973) and, lastly, 'Measuring the cognitive state of scientific disciplines' written by Cole, Cole and Dietrich in 1978.

These scholars employ an explorative and data-driven approach, which consists of putting forward certain variables that match their 'intuitive' ideas or folk theories of (the organization of) the sciences. The study by sociologist Norman Storer for example explains the data and approach in the following way: he extracts bibliometric information from 'two journals for each of ten fields of science, ranging from history to physics (…) one issue of each for the years 1926, 1936, 1946, 1956, and 1966, and [then we] counted things' (Storer, 1967: 80)

Storer's text is relatively short (9 pages) and is based on a talk he gave to medical librarians in 1966. Storer aims to offer analysis of 'the differences in the *qualities* of social relationships in the different sciences, or, perhaps, in the "atmospheres" or "moods" that characterize different fields of science.' (Storer, 1967: 75, italics in original) The quality that is operationalized empirically in this paper is 'hardness'. He starts with a discussion of connotations of hard, which implies 'tough', 'impenetrable' and 'impersonal. He then discusses the ways in which contributions to science are evaluated and argues that an evaluation always rests upon a relation between the contribution and what we already know. It is this relation that Storer argues is different in 'hard' versus 'soft' sciences. Harder sciences have a more tightly integrated body of knowledge, often because of the use of mathematics, which can measure the work's 'rigor'. Therefore in the hard sciences it is clearer whether a new contribution is right or wrong than in the soft sciences. Storer explains:

> I am proposing that the use of mathematics in a science provides a greater degree of precision in organizing its body of knowledge and, thus, a "tougher" set of criteria for the evaluation of new contributions. (…) I am suggesting that, through some faculty of folk-wisdom, we have hit upon a way to characterize different branches of science in terms of a continuum that measures essentially the tightness of integration of their various bodies of knowledge. (Storer, 1967: 78-79)

Interestingly, hardness is not only expressed as a characteristic of cognitive body of knowledge, but also implies different evaluation criteria and social relations between scientists. Storer argues that because in hard sciences evaluation criteria are more rigorous it is more easily apparent whether a contribution is right or wrong. Therefore, the risk of making a contribution is greater because colleagues can, more easily than in the soft sciences, 'hurt you' which for Storer implies that social relations will be more impersonal. Therefore, next to the use of mathematics as a measure of rigor, Storer interprets the use of initials, instead of full names, in reference lists as a measure of impersonality. Storer then tests these two measures (the data used are outlined above) and shows that they indeed fit with his idea of the hierarchy of the sciences. History journals used so few tables and had so few references with initials only that they were excluded from the analysis, leaving us with sociology, political science and psychology as soft sciences, botany, zoology and economics as medium hard, and physics, chemistry and biochemistry as the hard sciences.

For Storer, it is clear that this is indeed a hierarchy and not just a differentiation. In his conclusion he writes that he hopes the reader now understands more of 'the drive in the softer sciences to become more rigorous through the use of mathematics. This is not simply a desire to emulate the more successful sciences, but rather a desire for more effective grounds on which to organize the collective efforts of many scientists.' (Storer, 1967: 83)

Three years later, Price follows with the only text[1] in which he goes into detail about the humanities using bibliometric data. He attempts to differentiate between forms of scholarship, ranging from hard science to soft science and nonscience. Storer's article is a starting point for Price and is discussed as a background to his own work. While Storer operationalized hardness, Price aims to operationalize the extent to which knowledge is 'cumulative'. He explains:

> A now classical paper by Deutsch worked out in some detail the implications of a suggestion by Conant that the essential difference between the two modes of scholarship was that of 'cumulation' versus 'noncumulation'. It was seen that cumulation in this sense implies not merely growth, nor indeed growth at compound interest, but rather the existence of a tightly integrated structure for the sciences. Evidently the prototypes of the other side, identified as 'the humanities' grew (perhaps almost as fast), contained specialties and fashions just as science, but had something different from the integrated structure of cumulation. (Price, 1970: 4)

It is important to note here that for both Storer and Price there is an interrelation between the social relations of a domain of scholarship and the knowledge produced in the domain. Price explains:

> At this point it becomes evident that we cannot and should not artificially separate the matter of substantive content from that of social behavior. In order to deal with quantitative, highly ordered, rather certain findings, a special sort of social relation between participants is called for. (Price, 1970: 5)

The idea of a 'tightly integrated structure for the sciences' is therefore as much a social as a cognitive conceptualization of the difference between different forms of scholarship. It is developed empirically through a study of references in articles in different domains (including the humanities) of the Price index[2]. The index measures the percentage of references in articles to sources that are no more than 5 years old. The higher this score, the more the discipline has a distinguishable 'research front' and therefore a specific type of citation structure in terms of age of references. For humanistic scholarship Price finds that there is no distinguishable research front. Here he comes to his most explicit statement of how the humanities and the sciences have their own separate ways of doing things. Price explains, based on his empirical analysis of difference in Price index, that there are two metabolisms:

> It would seem that this index provides a good diagnostic for the extent to which a subject is attempting, so to speak, to grow from the skin rather than from the body. With a low index one has a humanistic type of metabolism in which the scholar has to digest all that has gone before, let it mature gently in the cellar of his wisdom, and then distill forth new words of wisdom about the same sorts of questions. In hard science the positiveness of the knowledge and its short term permanence enable (...) to emerge at the research front where interaction with one's peers is as important as the storehouse of conventional wisdom. Price, 1970: 15)

---

[1] This text was published three times: as a chapter in an edited book, as part of a conference proceeding of a world conference in sociology and as part of the extended edition (1986) of Price's celebrated book *Little science, big science*.
[2] This measure was introduced in Price, 1965 but this article did not empirically study differences between disciplines.

For both Storer and Price then reference behavior is an indicator for the social structure of scientific disciplines. Only particular, tightly integrated social structures can sustain knowledge production with a research front. After 'hardness' and 'cumulativeness' a third concept operationalized using bibliometric data was the concept of codification. Zuckerman and Merton, in their 1973 publication 'Age, aging and age structure in science', define codification as 'the consolidation of empirical knowledge into succinct and interdependent theoretical formulations.' This concept echoes both Storer's interest in the use of mathematics and Price's ideas on the 'tight integration of fields'. Indeed both are cited in Zuckerman and Merton's paper. However, codification is more explicitly cognitive than the measures developed by Storer and Price (see also Cozzens, 1985). Zuckerman and Merton argue that highly 'codified fields tend to obliterate the original versions of past contributions by incorporating their essentials in the new formulations' (p. 303). This makes citation analysis a useful measure as differences in codification are visible in the age of references. Drawing on Price's data, and adding some of their own, they use the Price index to measure codification. While this article contributed little empirically, it is important to note that the concept of codification is operationalized by using the same data as Price used for the concept of hardness.

The fourth publication in this period brings together the earlier work. Stephen Cole, Jonathan Cole and Lorraine Dietrich published their 'Measuring the cognitive state of disciplines' in 1978 as a chapter in an edited volume[3], reflecting on the 1972 Science Indicators report and including contributions of people like Eugene Garfield, Derek Price and John Ziman. The authors take a broader, more reflexive perspective on the development of bibliometric methods. Their analysis revolves around the notion of codification and its relation to the cumulative and progressive nature of different disciplines. They explain:

> In the work of Kuhn, and of Zuckerman and Merton, the suggestion is at least implicit that rapid incorporation of old work makes the discovery of new ideas more probable, since workers in these fields need not continually return to first principles, or develop their own logical framework. Rapid incorporation and a corresponding high immediacy of citations is an indicator of the extent to which a science is growing in a cumulative fashion. The extent to

---

[3] The edited volume entitled *Toward a metric of science* was edited by Yehuda Elkana, Joshua Lederberg, Robert Merton, Arnold Thackray and Harriet Zuckerman.

which recent work is utilized in current research may thus be seen as an indicator of the presence of conditions necessary for rapid scientific advance. (Cole, Cole & Dietrich, 1978: 222)

The authors employ the measure developed by Price, but also improve it by controlling for the total size of the literature in each discipline. Their findings are surprising. Between the natural and social sciences they find little difference in the 'immediacy effect' and, importantly, bigger differences between journals within the same field than between fields. In a separate analysis of two English literature journals, the only empirical engagement with the humanities, they find very low scores for the Price index. This leads them to conclude that:

> The immediacy effect may enable us to distinguish between a literature that is scientific and one that is not, even if it may not allow us to distinguish between highly codified and less codified scientific fields. (Cole, Cole & Dietrich, 1978: 226)

The four studies presented here show a common analytical strategy. The researchers collect sets of articles published in particular journals (assigned to a scientific domain) and analyze particular elements of these articles: the use of tables, the use of initials for first names of authors on reference lists, and the age of references. These elements are accorded significance as they are understood to offer a window on the social and/or cognitive structure of a community. Price for instance writes: 'A scholarly publication is not a piece of information but an expression of the state of a scholar or a group of scholars at a particular time (…)we can tell something about the relations amongst the people from the papers themselves.' (Price, 1970: 6)

In this way, references were used to operationalize three closely related concepts—scientificness, hardness and codification—with which the scholars in question wanted to compare the social and/or cognitive structure (or metabolism) of scientific domains. This research strategy was partially successful, but the issue of within-field variation could not be explained within this sociological framework. At the same time, by the end of the 1970s we see the rise of competing citation theories (see Cozzens, 1981; Wyatt et al., 2017) and a more general decline of the Mertonian sociology of science perspective of which all these authors were part (Luukkoonen, 1997). As we shall show, within bibliometrics this perspective remained relevant especially through the Price index.

The second configuration of the bibliometric system builds on the theoretical foundations of scholars such as Price. This literature emerged from library science, specifically collection management. In order to inform librarians, library and information scholars analyze the types of references included on reference lists of publications from different disciplines (the earliest examples are Gross & Gross, 1927, and Gross & Woodford, 1931). Most publications in this body of literature are descriptive in-depth case studies of the characteristics of references in a particular research area or discipline. Many of these studies are inspired by the Price index and include tables in which the percentage of references that are a maximum of 5 or 10 years old are compared between disciplines. While empirically these studies offer more depth, theoretically there is little development to be discerned.

The first of such studies that includes the humanities was written by Jones, Chapman and Woods (1972), who analyze references of articles in English history, distinguishing between medieval, early modern and late modern history. They write:

> Price suggested that the concentration of references to material published in the recent past, that is at the 'research front' of a subject, is the characteristic of a 'hard science', and that a 'soft science' or non-science has a high degree of archival literature. If a subject has approximately 42% of its references dated within the last five years, then it is a hard science; if it has between 42% and 21%, then it is a soft science; and if it has less than 21%, it is a non-science. By this standard, English history is clearly a non-science, a finding which will surprise nobody. (Jones, Chapman & Woods, 1972: 153)

The authors then present a table of percentages found in other studies for different disciplines. This research design is repeated in subsequent studies. Frost (1979) studies the function of citations in German literary research and makes a comparison, based on earlier research, with 'scientific' disciplines. Heinzkill (1980) studies a large sample of references in English literary research. While this study is descriptive for the most part, Heinzkill compares the age of references in his sample with earlier studies in the same way as Jones, Chapman and Woods (1972). Stern (1983) in a study of literary scholarship of specific authors and literary movements even reproduces the table produced by Jones, Chapman and Woods, and includes the results found in her study. Cullars (1985) analyzes references in monographs within American and British literary research and contextualizes these in comparison with other humanities disciplines, comparing his results with, amongst others, Heinzkill (1980), and Stern (1983). Other studies of a similar design include Budd (1986), Cullars (1992; 1995) and Thompson (2002). More recently, a similar approach has been used by Ardanuy, Urbano and Quintana (2009) to study Catalan literary studies, and by Hammarfelt (2012) to analyze references in Swedish literary studies.

Some studies that can be characterized as in-depth case studies of reference lists stand out, methodologically or theoretically, from the above mentioned group of studies. Brooks (1985) for instance studies citer motivations across disciplines by asking scholars about particular references made in their publications. He embeds his study theoretically in the citation debate and finds, empirically, significant differences between the 'humanities' subset and the 'science' subset of citer motivations. Heisey's (1988) study is designed to deductively test Kuhn's paradigm theory and Price's metabolism theory. He studies publications on the Dead Sea Scrolls in biblical archeology (as an example of a scientific discipline) and biblical criticism (as an example of a humanistic discipline) to empirically test the difference between science and humanities. His analysis confirms that references in biblical archeology are much younger and more often to journal articles, as he expected. Moreover, he shows that there is a concentration of archeology papers within the first years after the discovery of the Dead Sea Scrolls, which is not the case for biblical criticism. Heisey argues this shows that Price was right concerning the two ways in which literatures grow. In biblical archeology there was a research front that 'died' after a few years when the most important research puzzles were solved. In biblical criticism, on the other hand, such an effect is not observed as new perspectives continue to generate publications in the field.

Zwaan & Nederhof (1990) assert that previous research has, unjustly, argued that all humanities disciplines lack core journals (Cullars, 1985, makes this argument explicitly). In their study of theoretical linguistics they show that scholars do recognize a particular set of journals as core, and they show that the Price Index in theoretical linguistics is much higher than in other humanities disciplines. They conclude that compared to other disciplines theoretical linguistics does not fit with the 'humanistic stereotype'. Wiberley (2003) develops a bibliometric analysis of five types of scholarship, showing that in literary studies there are strong bibliometric differences between descriptive 'bibliographies', 'editing', 'historical studies', 'criticism' and 'theory'.

This second body of literature shares with the first an interest in the use of reference lists to describe and compare disciplines, and sometimes even to develop and/or test theories. They employ the Price index and other measures to analyze bibliometric differences and interpret these in terms of disciplinary differences. They are typically more interested in a diverse descriptive analysis of these reference lists, in terms of the type of publication cited, age, language and source type.

Conceptually it is especially the general opposition between science and humanities that 'survived', including the Price index as a valid but rough measure that can be used to compare disciplines. The studies engage with this opposition in different ways. Most of them take the opposition between science and humanities as a given (e.g. Jones, Chapman & Woods, 1972; Heinzkill, 1980; Stern, 1983; Cullars, 1985), empirically test it (Heisley, 1985) or refute it for specific humanities disciplines that behave more 'scientifically' than previously assumed (Zwaan & Nederhof, 1990). However, there is little independent theoretical development beyond this general opposition of scientific domains, which might be due to the different theoretical and practical background of many of these authors (being primarily situated in library and information science).

4. The third and fourth configuration of the bibliometric system: bibliometrics becomes a tool for research evaluation

By the end of the 1970s the field of bibliometrics grows into a more or less independent subfield of the larger field of library and information science. The commencement of the journal *Scientometrics* in 1978 is an important moment for scholars as it gives them a specialized outlet. The 1970s also sees the emergence of two new databases next to the SCI: the Social Science Citation Index (SSCI) and, more crucially for this article, the Arts & Humanities Citation Index (A&HCI)[4]. The community of bibliometric scholars grows and bibliometric analyses become increasingly important in the fields of science policy and science administration. This has a profound influence on the development of bibliometric scholarship. Research performance (e.g. Moed et al., 1985) becomes an increasingly important context in which bibliometric methods are developed.

Bibliometric publications become, within the context of research performance, more technical and policy-focused. This means that there is far less explicit theorizing regarding the cognitive or social structure of science and a greater concern for the practical usability of bibliometric methods. This influences the nature of bibliometric publications. Bibliometrics increasingly becomes an applied, data-driven field strongly embedded in science policy. The best ways to compare different units of analysis (faculties, universities, countries, disciplines) in a fair and legitimate way becomes the primary concern.

This shift has important implications for the conceptual framework invoked, types of bibliometric methods used, and the demands regarding data sources. First, concepts are no longer embedded in a sociological understanding of science (e.g. a Mertonian, interpretivist or constructivist framework) but in a science policy concern such as academic quality or internationalization. Second, the analysis of publication profiles becomes the most important bibliometric method of analysis (sometimes complemented with citation impact analysis). Third, data coverage becomes a crucial, and in the humanities highly problematic, issue that determines the extent to which a range of (citation-based) bibliometric indicators can be used at all.

In this part of our review we first discuss the origin and rise of 'internationality' as a new concept in bibliometric representations of the humanities as well as its relation to 'scientificness'. Second, we discuss the data sources, coverage issues and methodological problems that form the core of the division between the two configurations of the bibliometric system that we find in this period. In the third configuration of the bibliometric system, bibliometricians rely on the WoS databases supplemented with data extracted from annual reports and surveys, while in the fourth configuration bibliometricians rely on local, regional or national databases that provide full coverage of publications.

---

[4] In the remainder of the article we will use the term WoS databases to denote the three together.

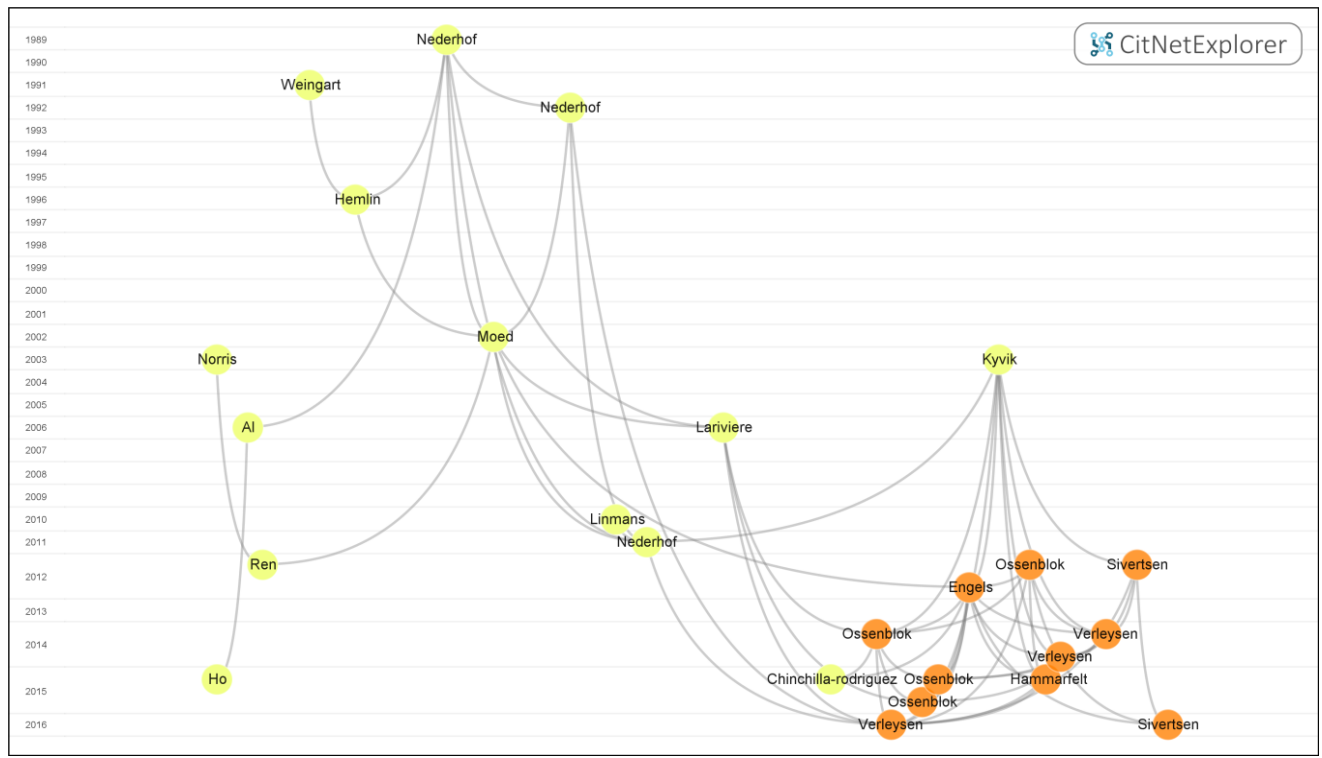

Figure 3: Visualization of the citation network of the third and fourth configuration of the bibliometric system

### 4.1 The conceptual shift towards internationalism

Two concepts are invoked throughout the bibliometric literature in this second period. The first is scientificness. This concept remains important but no longer as a character of the cognitive and social structure of a discipline. Rather, scientificness becomes defined first and foremost as a characteristic of the intended audience of publications. Publications aimed at a scientific audience are contrasted to publications aimed at a general audience. The second concept that emerges is internationality. As we shall show, the extent to which publications are 'international' becomes a new dividing line between the humanities and other domains and within the humanities itself. Furthermore, internationality and scientificness often intersect in a particular way. A local orientation is often assumed to be directed to a non-scientific audience, while an international orientation often invokes a scientific audience.

The research program of Nederhof and his colleagues in the Netherlands in the second half of the 1980s is the earliest example of this shift. At LISBON (which became CWTS in 1989), Nederhof, Zwaan, de Bruin and Dekker (Nederhof et al., 1988; Nederhof et al., 1989) studied the usefulness of bibliometric indicators for research evaluation in the humanities and social sciences. Nederhof et al. (1989) discuss Price (1970) and Cole (1983) and argue that the work of these scholars shows that the rate of scientific development in the humanities is slower. They then explain that there are two mechanisms responsible for this slower rate of development. First, humanities (and social science) researchers are argued to be more involved in 'enlightenment' of the non-scientific public. Second, these researchers are argued to publish more for a local (scientific) public, which also 'leads to a slower growth of knowledge'.

Nederhof et al. thus build on the work of Price and Cole and their notion of scientificness. However, they include the intended audience of publications, which is absent in the work of Price and Cole. Including non-journal publications and non-English publications is necessary to do justice to the research culture of the humanities in research evaluation, according to Nederhof and his colleagues. This aim is admirable and can be seen as a response to humanities' criticisms of the use of bibliometrics in research evaluation. It was especially the launch of the A&HCI, making citation impact analysis a possibility, that was feared in the humanities (e.g. Petrey, 1980). How Nederhof and colleagues proceed is however problematic.

Nederhof et al. (1989) study publication profiles in eight disciplines across the social sciences and humanities. Data was gathered from annual reports from universities in which particular departments (or parts of departments) could be connected to particular disciplines as well as from the A&HCI. The 'locality' of a publication plays a role in two ways. First, publications can be local because they are aimed at a general public (so-called enlightenment publications), and publications can be 'local' because they are written in a local language for a scientific public. The counter point for these two forms of locality is one form of internationality. Nederhof et al. take the language of publications, the nationality of the medium of publication, and coverage of articles in the WoS databases as variables in the operationalization of the notion of international. The authors conclude:

> The results suggest that a differentiated approach is called for concerning the use of bibliometric indicators. (…) The impact of departments differed greatly both within and between disciplines. For all fields, bibliometric indicators were potentially useful for monitoring international impact (…) [H]owever, ISI [WoS] citation data are not of much use in monitoring the national impact of Dutch scholars because of the low coverage of national journals. (Nederhof et al., 1989: 434)

The notion of 'national impact' is used in this quote to point to the citation impact of scientific publications written in Dutch. However, a different form of national impact is the impact of publications written for a general audience. Being of a local nature, independent of the scholarly content of the publication, becomes intertwined with the idea of non-scientificness and, vice versa, internationality becomes intertwined with scientificness. In this study we also see the emergence of the inclusion of journals/publications in the WoS database as a new indicator for internationalism. This is continuously reported in bibliometric studies as a meaningful measure (see also Sivertsen, 2016 for a critique of this particular phenomenon).

In a publication a few years later, Nederhof and Noyons (1992) 'assess the impact on various degrees of the dimension "local-cosmopolitan"' (Nederhof & Noyons, 1992: 250). Here, drawing on the A&HCI, they compare the citation impact of different general linguistics and general literature departments and the extent to which publications are cited by international or local authors. In the first German bibliometric study (Weingart et al., 1991) the authors discuss the publication profiles of different humanities disciplines in terms of types of publications (journals, edited volumes, monographs) and again language. They conclude that publications from German humanities scholars are mainly written in German and thus have a national orientation. For Hemlin (1996), studying Swedish archeology and English departments, publications in English and German are counted to measure the degree of internationalism of these departments. Bourke and Butler[5] (1996), who study Australian humanities and social sciences, do not account for language but do employ the dimension of internationalism, arguing that in WoS databases in the social sciences and humanities 'the under-representation of Australian and regional journals is an important reality.' (Bourke & Butler, 1996: 486)

Moed, Luwel & Nederhof (2002) argue that 'the extent to which research findings are communicated across national or cultural boundaries, is a relevant criterion of scholarly performance in all subdisciplines.' (2002: 513) While they explain that they are not arguing that all publications in English are better than those in Flemish, they do see publication in English as a valid indicator of international orientation and relate this to the scholarly importance of a publication. Here we see that, very explicitly, internationality of a publication (channel) becomes interlinked with the scholarly importance of a publication.

With the emergence of regional and national publication databases in Sweden, Norway and Flanders, a new data source becomes available for studies of humanities publication practices and profiles. The context of its emergence is discussed in more detail below. Here we want to note that also in the publications originating from this new data source a similar distinction is made between local and international/English publications. For instance, Sivertsen & Larsen (2012) distinguish between Scandinavian (domestic) and all other languages (international). Ossenblok, Engels & SIvertsen (2012) on the other hand distinguish between domestic (Flemish and Nordic), English and 'Other languages'. These studies routinely also include coverage in the WoS databases as a measure but, as is the case in previous studies, it is unclear what concept is operationalized through this particular measure.

As we have shown, there is a tendency in the studies that analyze publication profiles to include some form of the local-international as well as the coverage of publications in the WoS databases in their analysis. In some studies, publications that are aimed at a general audience and are most often written in what in those specific cases is a 'local' language are also included, but usually this is not the case. Especially in bibliometric studies that only analyze scholarly (peer-reviewed) publications what the language variable is meant to measure is not entirely clear.

Moed, Luwel & Nederhof (2002) suggest a relation between language and scientific importance. However, this is not always the case. As Sivertsen & Larsen (2012) suggest, it might make sense, from a scholarly perspective, to publish in a particular language to reach a scholarly audience. There is an overall tendency of methodological reductionism of research efforts to individual publications. It is very well possible that a publication written in a language with a small audience contains the same ideas as publications written in English for an audience that is equally scholarly but cannot be reached otherwise. Sivertsen (2016) recently showed that authors very often use different publication outlets at the same time. His findings suggest that it is highly problematic to assume that humanities publication practices will, eventually, become the same as the ideal-typical monolithic publication profile of the natural sciences. The peer-reviewed publication in a WoS-indexed journal is the norm with which humanities scholars are compared. The bibliometric field has not yet found ways to theoretically conceptualize divergence from this norm other than by alluding to the idea of diversity. However, what these findings suggest about the cognitive and social structure of the humanities has not been explored.

---

[5] This paper was excluded from the analysis because of the lack of citation relations with the other publications in the network

### 4.2 *Data sources and methods: coverage and the question of citation impact analysis in the humanities*

What unites the literature in the third and fourth configurations of the bibliometric system is their focus on publication profiles and their attention to the dimension of internationality as we outlined above. All these studies aim to show the diversity of publication types in the humanities. Within these publication profiles, language becomes the defining aspect that all studies use to differentiate between publications that have an international (scientific) audience and those that have a national (scientific or non-scientific) audience. However, the data sources that are used, and consequently the bibliometric methods that are employed, differ.

The studies in the third configuration of the bibliometric system employ a range of data sources and, very often, combine the WoS databases with a survey or data collected from annual reports of departments or universities. In some of these studies, namely those (co-)authored by Nederhof (Nederhof et al., 1989; Nederhof & Noyons; 1992, Nederhof 2011), Norris and Oppenheim (2003), and Linmans (2010), the possibility and legitimacy of citation impact analysis in the humanities is explored. The issue of whether citation impact analysis is legitimate comes to the foreground in this period because the context for bibliometrics studies has changed so dramatically compared to the 1960s and 1970s. As Moed, Luwel and Nederhof (2002) write in their introduction, bibliometricians aim to answer a particular question:

> How does one recognize a "good" scholar? (…) The approach adopted in this study can be defined as bibliometric. It aims at identifying characteristics of scholarly publications that can validly be assumed to reflect the "quality" or 'importance' of a scholar or a scholarly work. (Moed, Luwel & Nederhof, 2002: 499)

The question whether a particular bibliometric method can be used thus becomes a question of scientific *and* moral validity. The main point of dispute is whether citation impact analysis should be explored at all in the humanities. The primary issue here is the low coverage: both the limited coverage of humanities journals as well as the lack of coverage of books in all databases (e.g. Nederhof, 1989; Butler & Bourke, 1996; Hicks, 2004; Archambault et al., 2006, Linmans, 2010; Sivertsen, 2012), and differences in citation practices that might make the impact of humanities citations incomparable with those in other scientific domains. Doing citation impact studies in the humanities is systematically regarded as problematic including in those studies that experiment with it (Norris & Oppenheim, 2003; Ren & Gon, 2012, being exceptions).

There is an interesting difference between bibliometric schools of thought. Scholars that use the WoS databases (predominantly Dutch-based bibliometricians but also others) continue to experiment with citation impact analysis. Linmans (2010), for instance, develops new citation measures that fit with the on average lower amount of citations in the humanities. Nederhof (2011) argues that with a five-year citation window, citation impact analysis can be used in literature and linguistics, if supplemented with publication profiles to include books and other forms of output. However, outside the Netherlands (and outside bibliometrics) there is little systematic interest in the use of citation impact analysis in the humanities. Other studies that employ citation impact analysis are country case studies. In Norway, Denmark, Sweden and Flanders, the development of a partially performance-based funding system based on publication profiles has prevented the widespread use of citation impact analysis. The origin of these databases is described elsewhere (Sivertsen, 2006; Engels, Ossenblok & Spruyt, 2012; Sigl et al., 2017). What interests us is the type of bibliometric studies these databases might make possible.

These databases, and related university repositories that offer more or less complete publication records, provide new opportunities for the bibliometric research community. They allow for studies of publication profiles on larger scales (comparisons between institutions, disciplines and countries) and benefit from far better coverage than all earlier studies. This has led to a recent stream of new publications that use this data to analyze various aspects of publication practices in the humanities (and social sciences). These studies have had a predominantly descriptive character. An important research question has been whether the existence of performance-based funding mechanisms made possible by the databases has had an effect on publication practices (e.g. Hammerfelt & De Rijcke, 2016; Ossenblok, Engels, Sivertsen, 2012). Other studies analyze the role of edited books and the characteristics of book editors (Ossenblok, Verleysen & Engels, 2014; Ossenblok & Engels, 2015), Internationalization (Verleysen & Engels, 2014a; 2014b) and publication profiles of individual scholars (Verleysen & Weeren, 2016). These studies approach bibliometric data with novel methods, such as principal component analysis (Verleysen & Weeren, 2016) and the barycenter method (Verleysen & Engels, 2014a; 2014b).

5. <u>Is there a need for a renewed theoretization of bibliometrics?</u>

In this article we have analyzed how bibliometric representations of the humanities have been made in the history of bibliometric scholarship. We distinguish two periods of bibliometric scholarship, in which a total of 4 configurations of the bibliometric system are apparent. For each configuration we have analyzed the relation between concepts that are operationalized, bibliometric methods used, and data sources drawn upon.

The most significant change is the shifting conceptual hinterland in which bibliometric studies are embedded. In the first period, bibliometric studies are embedded in a theoretical framework derived from the sociology of science. This changes in the 1980s when bibliometric methods are increasingly used within the context of science policy and research performance studies. The conceptual hinterland shifts accordingly. In the first period the concept of 'scientificness' (and related concepts) is operationalized through the Price index, which is composed using the average age of references. Through the Price index bibliometricians make the claim that there is a relation between this average age of references and the extent to which the scientific domain or discipline has a research front, and therefore the extent to which it is scientific. Differences in Price indexes are shown to fit with an 'intuitive' hierarchy of scientific disciplines in which the humanities are characterized as a non-science because of the lack of research front.

In the second period, the concept of internationality is operationalized through the analysis of publication profiles of departments, disciplines and countries. Publications are ordered according to the language in which they have been written, the (inter)nationality of the publication venue, and the inclusion of the venue in the WoS databases. The scientific standard is an English-language article in a WoS-indexed peer reviewed journal, and bibliometricians explore the extent to which the humanities are comparable to this standard. Some scholars (e.g. Nederhof et al., 1989; Van Leeuwen, 2013) explore differences between the humanities and other scientific domains to make visible the diversity of humanities publication practices and to account for these in the analysis of research performance. They do not however employ a new conceptual framework in which research and publication practices in the humanities are conceptualized in their own right (however, see Sivertsen, 2016).

In both periods the humanities are a dominated 'other' to the ideal-typical research and publication practices of other scientific domains. It is therefore not surprising that bibliometricians have not been met with a lot of enthusiasm amongst humanities scholars (e.g. Dehue, 2000; Kiefer, 2014). There is indeed little reason to be supportive of bibliometric efforts from a humanities perspective.

What contemporary bibliometric studies lack is an explicit theoretical framework of the workings of science and the science system, in which bibliometric measures that are proposed can be understood as operationalizations of concepts from such a framework (see also Gläser & Laudel, 2016). 'Language of publication' and 'inclusion in the WoS databases' might serve as examples to illustrate this issue. In recent bibliometric studies both of these variables are used to measure the level of internationality but it is unclear what they mean beyond the performance measurement and research evaluation systems they might be part of. Similarly, there is a tendency for methodological reductionism in bibliometric studies in assuming each publication to be an autonomous research effort. Such methodological reductionism ignores the relations between (different types of) publications. Sivertsen (2016) recently showed that humanities scholars use a variety of publication forms at the same time. This implies that a research effort might be communicated through a monograph, a journal article and a newspaper article. As such, the research project might, from a research practice perspective, offer a more relevant unit of analysis. Similarly, groups of scholars work together in a variety of ways, often not resulting in co-authored articles but, for instance, in an edited volume.

Embedding bibliometric measures in a richer theoretical understanding of science and the science system is necessary for bibliometricians to resist pressures from science policy to deliver descriptive measures that have no clear relation to our understanding of research and publication practices. This also implies that bibliometricians involved in the humanities should not rely on a theoretical framework built around the natural sciences but rather develop an understanding of the way the humanities 'work', independent of other scientific domains. Despite all the shortcomings of the work of Price (1970) in representing the humanities, his argument that the humanities have a different metabolism of knowledge growth than the natural sciences offers an important starting point for bibliometrics of the humanities. For bibliometric analytical tools to become relevant for humanities scholars, it is critical that they become embedded in an empirically grounded understanding of the epistemic cultures of the humanities.